\documentclass[11pt,a4paper]{article}
\usepackage {graphicx}

\begin{document}

\title{Elastic Principal Graphs and Manifolds\\ and their Practical
Applications\thanks{Computing 75, 359--379 (2005), DOI:
10.1007/s00607-005-0122-6}}

\author{\textbf{A. Gorban}, Leicester, and \textbf{A. Zinovyev}, Paris}

\date{}
\maketitle

\begin{abstract}

Principal manifolds serve as useful tool for many practical
applications. These manifolds are defined as lines or surfaces
passing through ``the middle'' of data distribution. We propose an
algorithm for fast construction of grid approximations of
principal manifolds with given topology. It is based on analogy of
principal manifold and elastic membrane. First advantage of this
method is a form of the functional to be minimized which becomes
quadratic at the step of the vertices position refinement. This
makes the algorithm very effective, especially for parallel
implementations. Another advantage is that the same algorithmic
kernel is applied to construct principal manifolds of different
dimensions and topologies. We demonstrate how flexibility of the
approach allows numerous adaptive strategies like principal graph
constructing, etc.  The algorithm is implemented as a C++ package
\textit{elmap} and as a part of stand-alone data visualization
tool \textit{VidaExpert}, available on the web. We describe the
approach and provide several examples of its application with
speed performance characteristics. \end{abstract}



\section{Introduction}

Principal manifolds were introduced by Hastie and Stueltze in 1989
as lines or surfaces passing through ``the middle'' of the data
distribution \cite{HastieStuetzle89}. This intuitive definition
was supported by mathematical notion of self-consistency: every
point of the principal manifold is a conditional mean of all
points that are projected into this point. In the case of datasets
only one or zero data points are projected in a typical point of
the principal manifold, thus, one has to introduce smoothers that
become an essential part of the principal manifold construction
algorithms.

Since the pioneering work of Hastie, many modifications and
alternative definitions of principal manifolds have appeared in
the literature. Theoretically, existence of self-consistent
principal manifolds is not guaranteed for arbitrary probability
distributions. Many alternative definitions were introduced (see,
for example, \cite{KeglThesis99}) in  order to improve the
situation and to allow the construction of principal curves
(manifolds) for a distribution of points with several finite first
moments. A promising approach is based on analogy of principal
manifold and elastic membrane. The idea of using the elastic
energy functional for principal manifold construction in the
context of neural network methodology was proposed in mid 1990s
(see \cite{Gorban98,GorbanRossiev99} and bibliography there). This
idea was developed and tested on practical applications in
\cite{Gorban01,Gorban02,GorbanVisPreprint01,GorbanCHAOS01,
GorbanInfo00,GorbanNeuro02,Gorban03,ZinovyevBook00,GorbanOpSys03,Zinovyev02}.
Another computationally effective and robust algorithmic kernel
for principal curve construction, called the polygonal algorithm,
was proposed by K\'{e}gl et al. \cite{Kegl99}. A variant of this
strategy for constructing principal graphs was also formulated in
the context of the skeletonization of hand-written digits
\cite{Kegl02}. An interesting approach we would also like to
mention is the construction of principal manifolds in a piece-wise
manner by fitting unconnected line segments \cite{Verbeek00}.

Probably, most scientific and industrial applications of principal
manifold methodology were implemented using the Kohonen
Self-Organizing Maps (SOM) approach developed in the theory of
neural networks \cite{Kohonen82}. These applications are too
numerous to be mentioned here. We only mention that SOMs, indeed,
can provide principal manifold approximations (for example, see
\cite{Mulier95,Ritter92}) and are computationally effective. The
disadvantage of this approach is that it is entirely based on
heuristics; also it was shown that in the SOM strategy there does
not exist any objective function that is minimized by the training
process \cite{Erwin92}.

In this paper we introduce a computationally effective framework
for principal manifold construction. Our approach
\cite{GorbanCHAOS01,GorbanInfo00,ZinovyevBook00} combines ideas
developed in
\cite{Gorban98,GorbanRossiev99,GapsGRW,Gorban02,GapsGRW} with the
approach of K\'{e}gl \cite{KeglThesis99}, and takes some details
from the SOM approach as well. We use grid approximations to the
principal manifold, defining manifold in a finite number of
points. To describe elastic properties we utilize mesh of springs.
The topology of the manifold can be fixed or modified during the
process of construction.

Following metaphor of elasticity, we introduce two smoothness
penalty terms, which are quadratic at the vertex optimization
step. This allows using standard minimization of quadratic
functionals (i.e., solving a system of linear algebraic equations
with a sparse matrix), which is considerably more computationally
effective than gradient optimization of more complicated function,
introduced by K\'{e}gl.

Minimization of a positive definite quadratic functional can be
provided by the sequential one-dimensional minimization for every
space coordinate (cyclic). If for a set of coordinates $\{x_i\}_{i
\in J}$ terms $x_i x_j$ ($i,j \in J$, $i \neq j$) do not present
in the functional, then for these coordinates the functional can
be minimized independently. The quadratic functional we formulate
has a sparse structure, it gives us the possibility to use
parallel minimization that is expected to be particularly
effective in the case of multidimensional data.

Another feature of our approach is a universal and flexible way to
describe grid. A grid approximation to a principal manifold is
defined as a connected graph of nodes placed in data space and
having a ``natural'' node placement in a low-dimensional space.
The same algorithmic kernel is used to optimize the embedded graph
with respect to the dataset. Thus, the same algorithm, given an
initial definition of the grid, provides construction of principal
manifolds with different dimensions and topologies.

Our algorithm is implemented as a C++ package {\textit{elmap}}
\cite{Elmap} and as a stand-alone application
{\textit{VidaExpert}} for multidimensional data visualization
\cite{VidaExpert}. Some of the applications of the approach to the
data visualization were reported in series of works
\cite{Gorban01,Gorban02,GorbanVisPreprint01,GorbanCHAOS01,
GorbanInfo00,GorbanNeuro02,Gorban03,ZinovyevBook00,GorbanOpSys03,Zinovyev02}.

\section{Outline of the method}

We define an \textit{elastic net} as a connected unordered graph
$G(Y,$\textbf{\textit{E}}), where $Y=\{y^(i),\; i=1..p \}$ denotes
the collection of graph nodes, and $E=\{E^{(i )}, i=1..s$\} is the
collection of graph edges. We combine some of the incident edges
in pairs $R^{{\rm (}i{\rm )}~}$=~\{$E^{{\rm (}i{\rm )}} $ , $
E^{{\rm (}k{\rm )}}$\} and denote by
\textbf{\textit{R}}=\{$R^{{\rm (}i{\rm )}}, i=1..r$\} the
collection of \textit{elementary ribs}.

\begin{figure}
\centering{
\includegraphics[width=80mm, height=10mm]{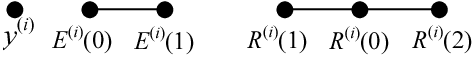}
} \caption{Node, edge and rib. \label{Fig1}}
\end{figure}

Every edge $E^{{\rm (}i{\rm )}}  $ has a beginning node $E^{{\rm
(}i{\rm )}}$(0) and an ending node $E^{{\rm (}i{\rm )}}$(1). An
elementary rib is a pair of incident edges. It has a beginning
node $R^{{\rm (}i{\rm )}}$(1), an ending node $R^{{\rm (}i{\rm )}}
$(2) and a central node $R^{{\rm (}i{\rm )}}$(0) (see
Fig.~\ref{Fig1}).

Introducing edges is equivalent to introducing connectivity on the graph;
this connectivity defines a topology of the principal manifold to be
constructed, along with its dimension. Ribs together with edges are used to
define a smoothness penalty function, defining in such a way a ``natural''
form of the graph. Edges connect pairs of nodes, ribs connect triples (or,
connect two nodes through another one).

\begin{figure}
\centering{
\includegraphics[width=120mm, height=63mm]{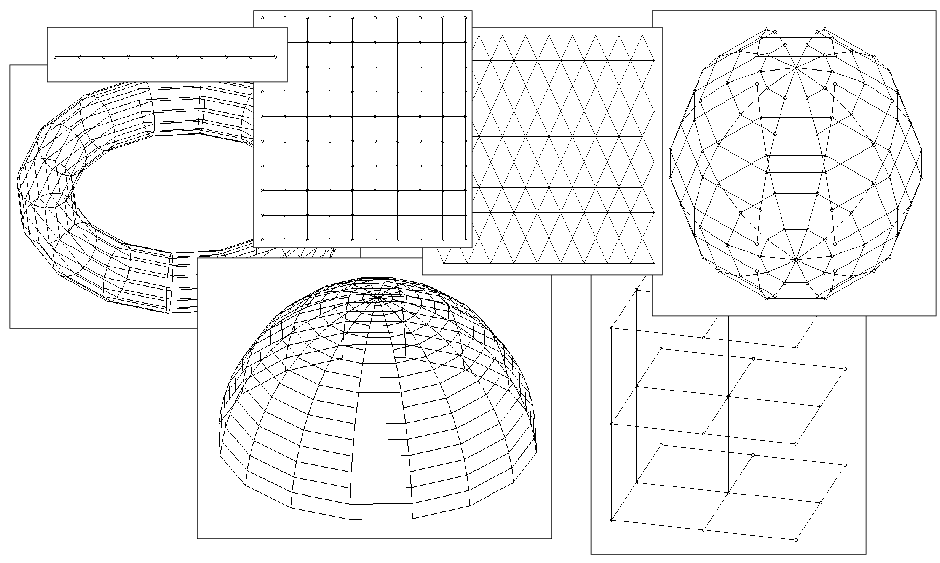}
} \caption{Elastic nets used in practice \label{Fig2}}
\end{figure}

Fig.~\ref{Fig2} illustrates some examples of the graphs
practically used. The first is a simple polyline, the second is a
planar rectangular grid, the third is a planar hexagonal grid and
the forth is a non-planar graph with nodes arranged on a sphere
(spherical grid), then a 3D cubical grid, torus and hemisphere.
Elementary ribs at these graphs are incident edges touching with a
blunt angle.

We underline here that the grids presented on Fig.~\ref{Fig2}
correspond to manifolds of different topology and dimension. The
grid embedded in data space is optimized with respect to the data
point positions.

In optimization criterion we use the standard mean squared
point-to-node distance as a main term, and two penalty terms,
which are useful to interpret in terms of physical elastic
properties of the grid.

For the graph $G$ we define the energy $U$ that includes energies
of every node, edge and rib:

\begin{equation}
U = U^{{\rm (}Y{\rm )}} + U^{{\rm (}E{\rm )}} + U^{{\rm (}R{\rm )}}.
\end{equation}

Let us divide the data points into subcollections $K^{{\rm (}i{\rm
)}}$, $i$~=~1\ldots $p$. The set $K^i$ contains the data points
for which the node $y^i$ is the closest one:

\[
K^{(i)} = \left\{x^{(j)}:{\left\| {x^{(j)} - y^{(i)}} \right\|}
\le {\left\| {x^{(j)} - y^{(m)}} \right\|},\; \mbox{for all} \; m
= 1,\ldots ,p\right\}.
\]

Let us  also assign a weight $w_{j}$ to every point. We define

\begin{equation}
\label{eq1}
U^{(Y)} = {\frac{{1}}{{{\sum\limits_{x^{(j)}} {w_{j}}} } }}{\sum\limits_{i =
1}^{p} {{\sum\limits_{x^{(j)} \in K^{(i)}} {w_{j}}} } } {\left\| {x^{(j)} -
y^{(i)}} \right\|}^{2},
\end{equation}

\begin{equation}
\label{eq2}
U^{(E)} = {\sum\limits_{i = 1}^{s} {\lambda _{i}}}  {\left\| {E^{(i)}(1) -
E^{(i)}(0)} \right\|}^{2},
\end{equation}

\begin{equation}
\label{eq3} U^{(R)} = {\sum\limits_{i = 1}^{r} {\mu _{i} {\left\|
{R^{(i)}(1) + R^{(i)}(0) - 2R^{(i)}(0)} \right\|}^{2}}} .
\end{equation}

\begin{figure}
\centering{
\includegraphics[width=90mm, height=50mm]{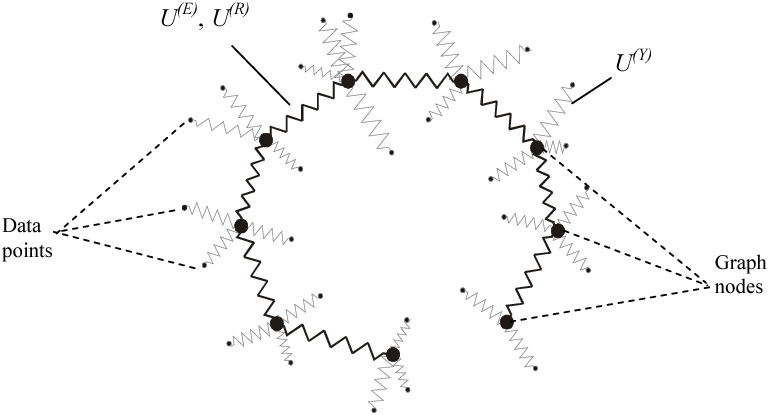}
} \caption{Energy of elastic net \label{Fig3}}
\end{figure}

The $U^{{\rm (}Y{\rm )}}$ term is the usual average weighted
square of distances between $y^{{\rm (}i{\rm )}}$ and data points
in $K^{{\rm (}i{\rm )}}$; $U^{{\rm (}E{\rm )}}$ is the analogue of
energy of elastic stretching and $U^{(R)}$ is the analogue of
energy of elastic bending of the net. We can imagine that every
node is connected by elastic bonds to the closest data points and
simultaneously to the adjacent nodes (see Fig.~\ref{Fig3}).

Values $\lambda _{i}$ and $\mu _{j}  $ are coefficients of
stretching elasticity of every edge $E^{{\rm (}i{\rm )}}$ and of
bending elasticity of every rib $R^{{\rm (}j{\rm )}}$. In the
simplest case we have

\[
\lambda _{1} = \lambda _{2} = ... = \lambda _{s} = \lambda (s),
\quad
\mu _{1} = \mu _{2} = ... = \mu _{r} = \mu (r).
\]

To obtain $\lambda (s)$ and $\mu (r)$ dependences we simplify the
task and consider the case of a regular, evenly stretched and
evenly bended grid. Let us consider a lattice of nodes of
``internal'' dimension $d$ ($d=1$ in the case of a polyline, $d=2$
in case of a rectangular grid, $d=3$ in the case of a cubical grid
and so on). Let the ``volume'' of the lattice be equal to $V$.
Then the edge length equals $\left( {V / s}
\right)^{{\raise0.7ex\hbox{${1}$} \!\mathord{\left/ {\vphantom
{{1}
{d}}}\right.\kern-\nulldelimiterspace}\!\lower0.7ex\hbox{${d}$}}}$.
Having in mind that for typical regular grids $r \approx s$, we
can calculate the smoothening parts of the functional:
$U^{(E)~}$\textit{$\sim $~}$\lambda s^{{\frac{{d - 2}}{{d}}}}$,
$U^{(R)~}$\textit{$\sim $~}$\mu r^{{\frac{{d - 2}}{{d}}}}$. Then
in the case where we want $U^{(R)}, U^{(E)}$ be independent on the
grid ``resolution'',

\begin{equation}
\label{eq4}
\lambda = \lambda _{0} s^{{\frac{{2 - d}}{{d}}}},
\quad
\mu = \mu _{0} r^{{\frac{{2 - d}}{{d}}}}
\end{equation}

\noindent where $\lambda _{ 0}$, $\mu _{ 0}$ are elasticity
parameters. This calculation is not applicable, of course, for the
general case of any graph. The dimension in this case can not be
easily defined and, in practical applications, the $\lambda _{i}$,
$\mu _{i} $ are often made different in different parts of a graph
according to some adaptation strategy (see below).

The elastic net approximates the cloud of data points and has
regular properties. Minimization of the $U^{{\rm (}Y{\rm )}}$ term
provides approximation, the $U^{{\rm (}E{\rm )}} $ penalizes the
total length (or, indirectly, ``square'', ``volume'', etc.) of the
grid and $U^{{\rm (}R{\rm )}} $ is a smoother term, preventing the
grid from folding and twisting.

In order to perform the vertex optimization step we derive the
system of algebraic linear equations to be solved. Let us consider
the situation when our collection of data points is already
separated in $K^{{\rm (}i{\rm )}}, \; i = 1\ldots p$.

Let us denote

\[
\Delta (x,y) = {\left\{ {\begin{array}{l}
 {1,x = y} \\
 {0,x \ne y,} \\
 \end{array}} \right.}
\]

\[
\Delta E^{ij} \equiv \Delta (E^{(i)}(0),y^{(j)}) - \Delta
(E^{(i)}(1),y^{(j)}),
\]

\[
\Delta R^{ij} \equiv \Delta (R^{(i)}(2),y^{(j)}) + \Delta
(R^{(i)}(1),y^{(j)}) - 2\Delta (R^{(i)}(0),y^{(j)}).
\]
That is, $\Delta E^{ij}=1$ if $y^j=E^{(i)}(0)$, $\Delta E^{ij}=-1$
if $y^j=E^{(i)}(0)$, and $\Delta E^{ij}=0$ for all other $y^j$;
$\Delta R^{ij}=1$ if $y^j=R^{(i)}(1)$ or $y^j=R^{(i)}(2)$, $\Delta
R^{ij}=-2$ if $y^j=R^{(i)}(0)$, and $\Delta R^{ij}=0$ for all
other $y^j$. After a short calculation we obtain the system of $p$
\textit{linear} equations to find new positions of nodes in
multidimensional space \{$y^{i}, i=$1\textit{\ldots p}\}:

\[
{\sum\limits_{k = 1}^{p} {a_{jk} y^{(k)} =
{\frac{{1}}{{{\sum\limits_{x^{(i)}} {w_{i}}} }
}}{\sum\limits_{x^{(i)} \in K_{j}} ^{} {w_{i} x^{(i)}}}} },
\]

where

\begin{equation}
 a_{jk} = {\frac{{n_{j} \delta _{jk}}} {{{\sum\limits_{x^{(i)}} {w_{i}}} } }}
+ e_{jk} + r_{jk} , \quad j = 1\ldots p ,
\end{equation}

\[
\delta _{jk} = {\left\{ {\begin{array}{l}
 {1,{\mathop {}\limits^{}} i = j} \\
 {0,{\mathop {}\limits_{}} i \ne j} \\
 \end{array}} \right.}
\]

\noindent and $n_{j}  =  {\sum\limits_{x^{(i)} \in K^{(j)}}^{}
{w_{i}}}  $, $e_{jk} = {\sum\limits_{i = 1}^{s} {\lambda _{i} \Delta
E^{ij}\Delta E^{ik}} }$, $r_{jk} = {\sum\limits_{i = 1}^{r} {\mu
_{i} \Delta R^{ij}\Delta R^{ik}} }$. The values of $e_{jk}$ and
$r_{jk}$ depend only on the structure of the grid. If the structure
does not change then they are constant. Thus only the diagonal
elements of the matrix (6) depend on the data set. The $a$ matrix
has sparse structure for a typical grid used in practice. In the
Appendix we define this structure, giving an algorithm for
calculating only non-zero elements of the matrix.

To minimize the energy of the graph $U $ we use the following
two-step iterative algorithm:
\begin{enumerate}
\item{Initialize the grid of nodes in data space.}
\item{Given the nodes placement, separate the collection of data
points into subcollections $K^{{\rm (}i{\rm )}}$, $i$ = 1\ldots
$p$.}
\item{Given this separation, minimize the graph energy $U$ and
calculate new positions of nodes.}
\item{Go back to step 2.}
\end{enumerate}
It is evident that this algorithm converges to a final placement
of nodes of the grid (energy $U$ is a non-decreasing value, and
the number of divisions of data points into $K^{{\rm (}i{\rm )}}$
is finite). Moreover, theoretically the number of iterations of
the algorithm before converging is finite. In practice this number
may be too large; therefore we interrupt the process of
minimization if change of $U  $ becomes less than a small value
$\varepsilon $ or after a fixed number of iterations.

\section{Optimization strategies}

\begin{figure}
\centering{
\includegraphics[width=130mm, height=30mm]{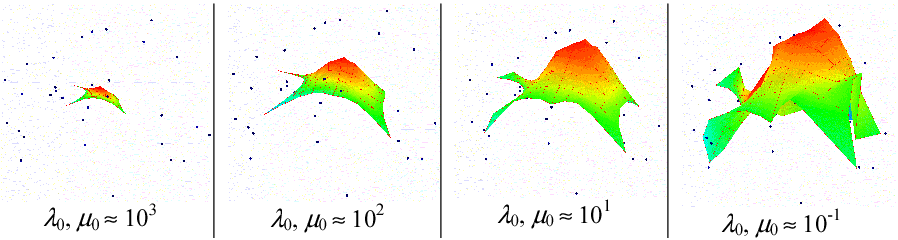}
} \caption{Training elastic net in several epochs (softening)
\label{Fig4}}
\end{figure}

We can guarantee that the algorithm described at the end of the
previous section leads to a local minima of the functional only.
Obtaining a solution close to the global minimum can be a
non-trivial task, especially in case where the initial position of
the grid is very different from the expected (or unknown) optimal
solution. In many practical situations the ``softening'' strategy
can be used to obtain solutions with low energy levels robustly.
This strategy starts with ``rigid'' grids (small length, small
bending and large $\lambda $, $\mu $ coefficients) at the
beginning of the learning process and finishes with soft grids
(small $\lambda $, $\mu $ values), Fig.~\ref{Fig4}. Thus, the
training goes in several epochs, each epoch with its own grid
rigidness. The process of ``softening'' is one of numerous
heuristics  that pretend to find the global minimum of energy $U$
or rather close configuration.

Nevertheless, for some artificial distributions (like spiral point
distribution, used as a test in many papers on principal curves
construction) ``softening'' starting from any linear configuration
of nodes does not lead to the expected solution. In this case,
adaptive strategies, like ``growing curve'' (analogue of what was
used by K\'{e}gl in his polygonal algorithm \cite{Kegl99} or
``growing surface'' can be used to obtain suitable configuration
of nodes. This configuration does not have to be optimal, in the
adaptation process one can still use the grids more rigid than it
is needed for good approximation (thus, providing more robust ways
of doing this), finishing the optimization at the next stage with
a softer grid (see \textit{spiral} example in the examples
section).

\section{Adaptive strategies}

The method described above allows us to construct different
adaptive strategies by playing with a) individual $\lambda _{i}$
and $\mu _{j}  $ weights; b) the grid connection topology; c) the
number of nodes.

This is a way of extending the approach significantly making it
suitable for practical applications. The {\textit{elmap}} package
with implementation of the method described above supports several
adaptive strategies that will be described in this section.

First of all, let us define a basic operation on the grid, which
allows inserting new nodes. Let us denote by \textbf{N},
\textbf{S}, \textbf{R} the sets of all nodes, edges and ribs
respectively. Let us denote by \textbf{C}$(i)$ the set of all
nodes which are connected to the $i$th node by an edge. If one has
to insert a new node in the middle of an edge $I$, connecting two
nodes $k$ and $l$, then the following operations have to be
accomplished:
\begin{enumerate}
\item{Delete from \textbf{R} those ribs which contain node $k$ or
node $l$;}
\item{Delete the edge $I  $ from \textbf{S};}
\item{Put a new node $m  $ in \textbf{N};}
\item{Put in \textbf{S} two new edges connecting $k$ and $m$, $m$ and
$l$;}
\item{Put in \textbf{R} new ribs, connecting $m$, $k$ and all
$i \in $\textbf{C}($k)$, and $m$, $l$ and all $i \in
$\textbf{C}($l)$.}
\end{enumerate}
At steps 4, 5 one has to assign new weights to the edges and ribs.
This choice depends on the task to be solved. If one constructs a
``growing'' grid, then these weights must be chosen the same as
they were at the deleted ones. If one constructs a refinement of
an already constructed grid, one must choose these weights to be
twice bigger than they were at the deleted ones.

The \textit{grow-type strategy} is applicable mainly to grids with
planar topology (linear, rectangular, cubic grids). It consists of
an iterative determining of those grid parts, which have the
largest ``load'' and doubling the number of nodes in this part of
the grid. The load can be defined in different ways. One natural
way is to calculate the number of points that are projected onto
the nodes. For linear grids the grow-type strategy consists of
\begin{enumerate}
\item{Initializing the grid; it must contain at least two nodes and
one edge;}
\item{Determining the edge which has the largest load, by
summing the number of data points (or the sum of their weights)
projected to both ends of every edge;}
\item{Inserting a new node in
the middle of the edge, following the operations described above;}
\item{Optimizing the positions of the nodes.}
\end{enumerate}
One stops this process usually when a certain number of nodes in
the grid is reached (see, for example, \cite{Kegl00}). This number
is connected with the total amount of points. In the
{\textit{elmap}} package this is an explicit parameter of the
method, allowing the user to implement his own stopping criterion.
Because of this stopping condition the computational complexity is
not proportional to the number of data points and, for example,
grows like $n^{{\rm 5}{\rm /} {\rm 3}}  $ in the case of the
Polygonal Line algorithm. Another form of the stopping condition
is when the mean-square error (MSE) does not change more than a
small number $\varepsilon $ after several insertion/optimization
operations.

We should mention here also {\it growing lump} and {\it growing
flag} strategies used in physical and chemical applications
\cite{Grids,CMIM}. In growing lump strategy we add new nods
uniformly at the boundary of the grid using a linear extrapolation
of the grid embedding. Then the optimization step follows, and,
after that, again the step of growing could be done.

For the invariant flag one uses sufficiently regular grids, in
which many points are situated on the coordinate lines, planes,
etc. First, we build a one-dimensional grid (as a one-dimensional
growing lump, for example). Then we add a new coordinate and start
growing in new direction by adding lines. After that, we can add
the third coordinate, and so on.

The \textit{break}-type adaptive strategy changes individual rib
weights in order to adapt the grid to those regions of data space
where the ``curvature'' of data distribution has a break or is
very different from the average. It is particular useful in
applications of principal curves for contour extraction (see
Fig.~\ref{Fig7}). For this purpose the following steps are
performed:
\begin{enumerate}
\item{Collect statistics for the distances from every node $i$ to
the mean point of the datapoints that are projected into this
node: $$r_{j} = {\left\| {y_{j} - \left(\sum\limits_{x^{(i)} \in
K_{j}} {w_{i}}\right)^{-1} {\sum\limits_{x^{(i)} \in K_{j}} ^{}
{w_{i} x^{(i)}}}} \right\|}.$$}
\item{Calculate mean and standard deviation for some power of
{\textit{r}} : $m = \overline {r^{\alpha}}  $, $s = \sigma
_{r^{\alpha}}  $; where $\alpha>1$ is a parameter which in our
experiments is chosen to be 4.}
\item{Determine those nodes for
which $r_{j} > m+\beta s$, where $\beta > 0$ is another parameter,
equal 2 in our experiments.}
\item{For every node $k$ determined
at the previous step one finds those ribs that have $k$ as their
central point and change their weight for $\mu _{j}^{(new)} = \mu
_{j}^{(old)} \cdot {\frac{{m}}{{r_{j}^{\alpha}} } }$.}
\item{Optimize the node positions.}
\item{Repeat this process a certain number of times.}
\end{enumerate}
\textit{Principal graph} strategy, implemented in the
{\textit{elmap}} package allows performing clustering of
curvilinear data features along principal curves. Two example
applications of this approach are satellite image analysis
\cite{Banfield92} or hand-written symbol skeletonization
\cite{Kegl02} (see also Fig.~\ref{Fig8},\ref{Fig9}. First, notice
that the grid we constructed does not have to be a connected
graph. The system matrix (6) is not singular if for every
connected component of the graph there are data points that are
projected onto one of its nodes. This allows using the same
algorithmic kernel to optimize node positions of unconnected
graph. Notice also that if the sets of edges and ribs are empty,
then this algorithm acts exactly like standard K-means clustering.

To construct a ``skeleton'' for two-dimensional point distribution,
we apply a variant of local linear principal component analysis
first, then connect local components into several connected parts
and optimize the node positions after. This procedure is robust and
efficient in applications to clustering along curvilinear features
and it was implemented as a part of {\textit{elmap}} package. The
following steps are performed:
\begin{enumerate}
\item{Make a ``grid'' from a number of unconnected nodes (sets of edges
and ribs are empty at this stage). Optimize the node positions
(i.e., do $K$-means clustering). The number of nodes is chosen to
be a certain proportion of the number of data points. In our
experiments we used 5\% of the total number of data points. At
every iteration of the $K$-means algorithm, the ``empty'' nodes
(those for which there is no data point having this node as this
closest one) change their position randomly. After a certain
number of K-means iterations, empty nodes (or nodes with only one
datapoint as well) are removed from the set of all nodes.}
\item{For
every node of the grid in position $y^{i}$, the local first
principal direction is calculated. By local we mean that the
principal direction is calculated inside the cluster of datapoints
corresponding to the node $i$. Then this node is substituted by
two new nodes in positions $y^{(new1)}= y^{i}+\alpha
s\textbf{\textit{n}}, y^{ (new2)}= y^{i} - \alpha s
\textbf{\textit{n}}$, where \textbf{\textit{n}} is the unit vector
in the principal direction, $s$ is the standard deviation of data
points belonging to the node $i$, $\alpha $ is a parameter, which
can be taken to be 1. These two nodes are connected by an edge
(see Fig.~\ref{Fig9}b).}
\item{A collection of edges and ribs is generated,
following this simple rule: every node is connected to the node
which is closest to this node but not already connected at the
step 2, and every such connection generates two ribs, consisting
of a new edge and one of the edges made at step 2.}
\item{Weights
of the ribs are calculated. A rib is assigned a weight equal to
$\vert $cos($\alpha )\vert $, where $\alpha $ is an intersection
angle of two edges contained in this rib, if $\alpha   \ge
\frac{\pi }{2} $. Otherwise it is zero (or, equally, the rib is
eliminated).}
\item{The node positions are optimized.}
\end{enumerate}
One possible way to improve the resulting graph further is to
apply graph simplification rules, analogously to how it was done
in \cite{Kegl02}. The idea of this algorithm is close to the
$k$-segments algorithm of Verbeek \cite{Verbeek00} and, indeed,
one possible option is to use $k$-segment clustering instead of
K-means clustering on the first step of the algorithm.

The adaptive strategies: ``grow'', ``break'' and the principal
graphs can be combined and applied one after another. For example,
the principal graph strategy can be followed by break-type weight
adaptation or by grow-type grid adaptation.

\section{Projecting}

In the process of the grid construction we use projection of data
into the closest node. This allows us to improve the speed at the
data projection step without loosing too much when the grid
resolution is good enough. The effect of an estimation bias,
connected with this type of projection, was observed in
\cite{KeglThesis99}. In our approach the bias is indirectly
reduced by utilizing the $U^{(E)}$ smoother term that makes the
grid almost isometric (having the same form, the grid will have
lesser energy with equal edge lengths). For presentation of data
points or for data compression, other projectors can be applied. A
natural way to do it is to introduce a set of simplexes on the
grid (line segments for one-dimensional grids, triangles for
two-dimensional grids, and tetrahedrons for the 3D grids). Then
one performs orthogonal projection onto this set. In order to not
calculate all distances to all simplexes, one can apply a
simplified version of the projector: find the closest node of the
grid and then consider only those simplexes that contain this
node. This type of projection is used in the {\textit{elmap}}
package and demonstrated by the example on Fig.~\ref{Fig9}.

Since the grid has penalty on its length (and, for higher
dimensions, indirectly, area, volume), the result of the
optimization procedure is a bounded manifold, embedded in the
cloud of data points. Because of this, if the penalty coefficient
is big, many points can have projection on the boundary of the
manifold. This can be undesirable, for example, in data
visualization applications. To avoid this effect, we introduced in
the {\textit{elmap}} package the possibility to make a linear
extrapolation of the bounded rectangular manifold (extending it by
continuity in different directions). Other, more complicated
extrapolations can be performed as well, like using Carleman's
formulas (see
\cite{Aizenberg93,Grids,CMIM,Gorban02,GapsGRW,GapsDGRKM}).

\section{Principal manifold as elastic membrane}

Let us discuss in more detail the central idea of this paper:
using metaphor of elastic membrane in principal manifold
construction algorithm. The system represented on Fig.~\ref{Fig3}
can be modeled as elastic membrane with external forces applied to
the nodes. In this section we consider the question of
correspondence between our spring network system and realistic
physical systems (evidently, we make comparison in 3D).

Spring meshes are widely used to create physical models of elastic
media (for example, \cite{Born54}). The advantages, comparing with
the continuous approaches like Finite Elements Method (FEM), are
evident: computational speed, flexibility, possibility to solve
the inverse elasticity problem easily \cite{VanGelder97}.

Modeling elastic media by spring networks has a number of
applications in computer graphics, where, for example, there is a
need to create realistic models of soft tissues (human skin, as an
example). In \cite{VanGelder97} it was shown that it is not
generally possible to model elastic behavior of a membrane using
spring meshes with simple scalar springs. In \cite{Xie02} the
authors introduced complex system of penalizing terms to take into
account angles between scalar springs as well as shear elasticity
terms. This allowed to improve the results of modeling and develop
applications in subdivision surface design.

In a recent paper \cite{Gusev04} it was demonstrated that there is
an exact correspondence between the FEM approach and spring
networks where elastic behavior of every spring is defined by
$6\times 6$ matrix

\[
K^{S} = \left( {{\begin{array}{*{20}c}
 {k^{s}} \hfill & { - k^{s}} \hfill \\
 { - k^{sT}} \hfill & {k^{sT}} \hfill \\
\end{array}}}  \right)
\]

\noindent where $k^{s}$ is a $3\times 3$ matrix describing the
elastic behavior of spring $s  $ with one of the two ends fixed.
In particular, to model 2D-elastic membrane by a regular
close-packed triangular lattice spring model, one takes the
springs with the following stiffness matrix (in the coordinate
frame where the spring is oriented along the $x$-axis)

\begin{equation}
\label{eq5}
k^{s} = {\frac{{1}}{{2\sqrt {3}}} }\left( {{\begin{array}{*{20}c}
 {3{\lambda} ' + 5{\mu} '} \hfill & {0} \hfill \\
 {0} \hfill & {{\mu} ' - {\lambda} '} \hfill \\
\end{array}}}  \right),
\end{equation}

\noindent where ${\lambda} '$ and ${\mu} '$ are Lam\'{e} constants
of the initial membrane. Simple scalar springs can be utilized only
in the particular case ${\lambda} ' = {\mu} '. $

Let us slightly reformulate our problem to make it more close to
the standard notations in the elasticity theory. We introduce the
$m \times p$-dimensional vector of displacements, stacking all
coordinates for every node:

\[
u = \{u_{1}^{(1)} ;u_{2}^{(1)} ;..;u_{m}^{(1)} ;...;u_{1}^{(p)}
;u_{2}^{(p)} ;..;u_{m}^{(p)} \}^{T}\]

\noindent where $m$ is dimension, $p$ is the number of nodes,
$u_{i}^{(k)}  $ is the $i$th component of the $k$th node
displacement. The absolute positions of nodes are $y_{}^{(k)} =
\tilde {y}_{}^{(k)} + u_{}^{(k)} $, where $\tilde {y}_{}^{(k)} $
are equilibrium (relaxed) positions. Then our minimization problem
can be stated in the following generalized form:

\begin{equation}
\label{eq6}
\quad
u^{T}Eu + D(u;x) \to \min ,
\end{equation}

\noindent where $E  $ is a symmetric $(m \times p) \times (m
\times p)$ element stiffness matrix. This matrix reflects elastic
properties of the spring network and has the following properties:
1) it is sparse; 2) it is invariant with respect to translations
of the whole system (as a result, for any band of $m$ consecutive
rows corresponding to a given node $k$, the sum of the m$m \times
m$ off-diagonal blocks should always be equaled to the
corresponding diagonal block taken with the opposite sign). The
$D(u;x)$ term describes how well the set of data $x  $ is
approximated by the spring network with the node displacement
vector $u$. It can be interpreted as the energy of external forces
applied to the nodes of the system. To minimize (\ref{eq6}) we
solve the problem of finding equilibrium between elastic internal
forces of the system (defined by $E$) and external forces:

\begin{equation}
\label{eq7}
Eu = f,\;\;\;f = - {\frac{{1}}{{2}}}{\frac{{\partial}} {{\partial
u}}}D(u;x).
\end{equation}

In the method introduced above, we propose to assemble the matrix
$E$ with use of simple scalar springs plus ribs to introduce bending
elasticity. The matrix is assembled very similar to how it is
described in the Appendix. There is one important point: the springs
(edges) have zero rest lengths, it means that equilibrium node
positions are all in zero: $\tilde {y}_{}^{(k)} = 0,\;k = 1..p$. The
system behavior then can be described as ``super-elastic''. From the
point of view of data analysis it means that we do not impose any
pre-defined shape on the data cloud structure.

\begin{figure}
\centering{
\includegraphics[width=75mm, height=40mm]{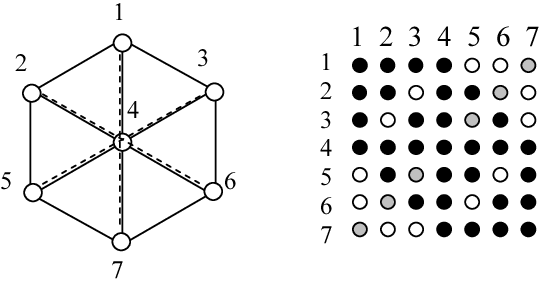}
} \caption{The stiffness matrix structure for one particular
elastic graph. Dashed lines denote ribs. Black circles correspond
to local connections of nodes. Gray circles correspond to
non-local connections (inside ribs), through one node
\label{Fig5}}
\end{figure}

Let us look at the structure of $E$ for a simple configuration of
nodes, see Fig.~\ref{Fig5}. Edges give local connections, whereas
the ribs produce terms that describe connection of two nodes
through another (in a rib two ending nodes are connected through
the central one). These non-local connections are marked on
Fig.~\ref{Fig5} by gray circles. This observation tells that
generally our stiffness matrix differs in its structure from the
one used in the FEM approach (where all connections are local).
The same is true for the system of terms used in \cite{Xie02}: for
example, the term penalizing angle deviations introduces non-local
connections in the corresponding stiffness matrix.

For $D(u$;$x)$ we use the usual mean square distance measure, see (\ref{eq1}): $D(u;x) =
U^{(Y)}$. The force applied to the $j$th node equals

\begin{equation}
\label{eq8}
f_{j} = {\frac{{n^{(j)}}}{{N}}}\left( {\tilde {x}^{(j)} - u^{(j)}}
\right)\;\;,
\end{equation}

\noindent where

\[
\tilde {x}^{(j)} = {\frac{{{\sum\limits_{x^{(i)} \in K^{(j)}}^{}
{w_{i} x^{(i)}}}} }{{n^{(j)}}}},\;\;n^{(j)} = {\sum\limits_{x^{(i)}
\in K^{(j)}} {w_{i}}}  ,\;\;\;\;N = {\sum\limits_{x^{(i)}} {w_{i}}}
\]

It is proportional to the vector connecting the $j$th node and the
weighted average $\tilde {x}^{(j)} $ of the data points in
$K^{{\rm (}j{\rm )}}_{ } $ (i.e., the average of the points that
surround the $j$th node: see (\ref{eq1}) for definition of
$K^{{\rm (}j{\rm )}})$. The proportionality factor is simply the
relative size of $K^{{\rm (}j{\rm )}}$. The linear structure of
(\ref{eq8}) allows to move $u$ in the left part of the equation
(\ref{eq8}). Thus the problem is linear.

Now let us show how we can benefit from the definition (\ref{eq7})
of the problem. First, we can introduce a pre-defined equilibrium
shape of the manifold: this initial shape will be elastically
deformed to fit the data. This approach corresponds to introducing
a model into the data. After we assemble a physically realistic
stiffness matrix $E  $ constructed following the recipe from
\cite{Gusev04}. In a particular but very practical case of a
regular close-packed triangular lattice spring model we assemble
$E$ using individual spring matrices in the form (\ref{eq5}).

Secondly, one can try to change the form (\ref{eq8}) of the external forces applied
to the system. In this way one can utilize other, more sophisticated
approximation measures: for example, taking the outliers into account.

Third, in three-dimensional applications one can benefit from
existing solvers for finding equilibrium form of elastic
membranes. They can be utilized to solve the problems analogous to
the one shown on Fig.~\ref{Fig7}. For multidimensional data point
distributions one has to adapt the engines, but this adaptation is
mostly formal.

Finally, there is a possibility of a hybrid approach: we utilize first
``super-elastic'' energy functional (1) to find the initial approximation.
Then we ``fix'' the result and define it as the equilibrium. After we
utilize physical elastic functional to find elastic deformation of the
equilibrium form to fit the data.

\section{Examples}

\begin{figure}
\centering{ a) spiral \includegraphics[width=45mm,
height=45mm]{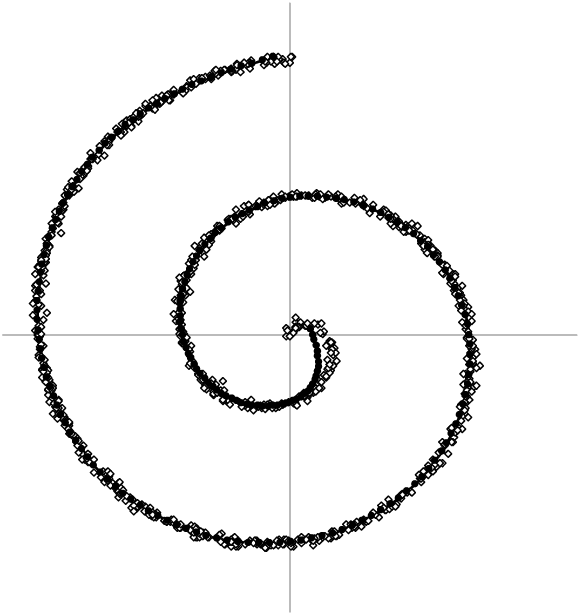} b) large \includegraphics[width=45mm,
height=45mm]{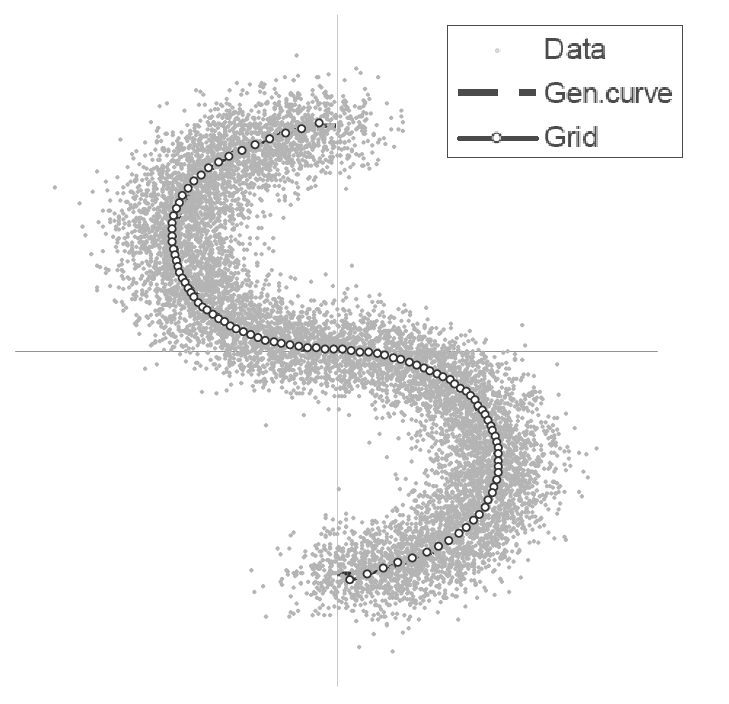} } \caption{Two-dimensional examples of
principal curves construction \label{Fig6}}
\end{figure}

On Fig.~\ref{Fig6} we present two examples of 2D-datasets provided
by K\'{e}gl\footnote{http://www.iro.umontreal.ca/ $ \sim
 $ Kegl/research/pcurves/implementations/Samples/}.

The first dataset called \textit{spiral} is one of the standard in
the principal curve literature ways to show that one's approach has
better performance than the initial algorithm provided by Hastie and
Stuelze. As we have already mentioned, this is a bad case for
optimization strategies, which start from linear distribution of
nodes and try to optimize all the nodes together in one loop. But
the adaptive ``growing curve'' strategy, though being by order of
magnitude slower than the ``softening'', finds the solution quite
stably, with exception for the region in the neighborhood of zero,
where the spiral has very different (comparing to the average)
curvature.

Second dataset, called \textit{large} is a simple case, despite
the fact that it has comparatively large sample size (10000
points). The nature of this simplicity lies in the fact that the
initial first principal component based approximation is already
effective; the distribution is in fact \textit{quasilinear}, since
the principal curve can be unambiguously orthogonally projected
onto a line. On Fig.~\ref{Fig6}b it is shown that the generating
curve, which was used to generate this dataset, has been
discovered almost perfectly and in a very short time. To give the
idea of speed, we mention that in the case of the simplest
optimization (one epoch with fixed grid rigidness, which is
suitable in the case of a good initial approximation) the
algorithm we described gives the principal curve, approximated by
100 nodes in less than 0.5 seconds on a computer with an Athlon
1800 MHz processor. Application of a softening strategy with 4
epochs gives the principal curve in approximately 1.5 seconds on
the same computer.

\begin{figure}
 \includegraphics[width=40mm, height=40mm]{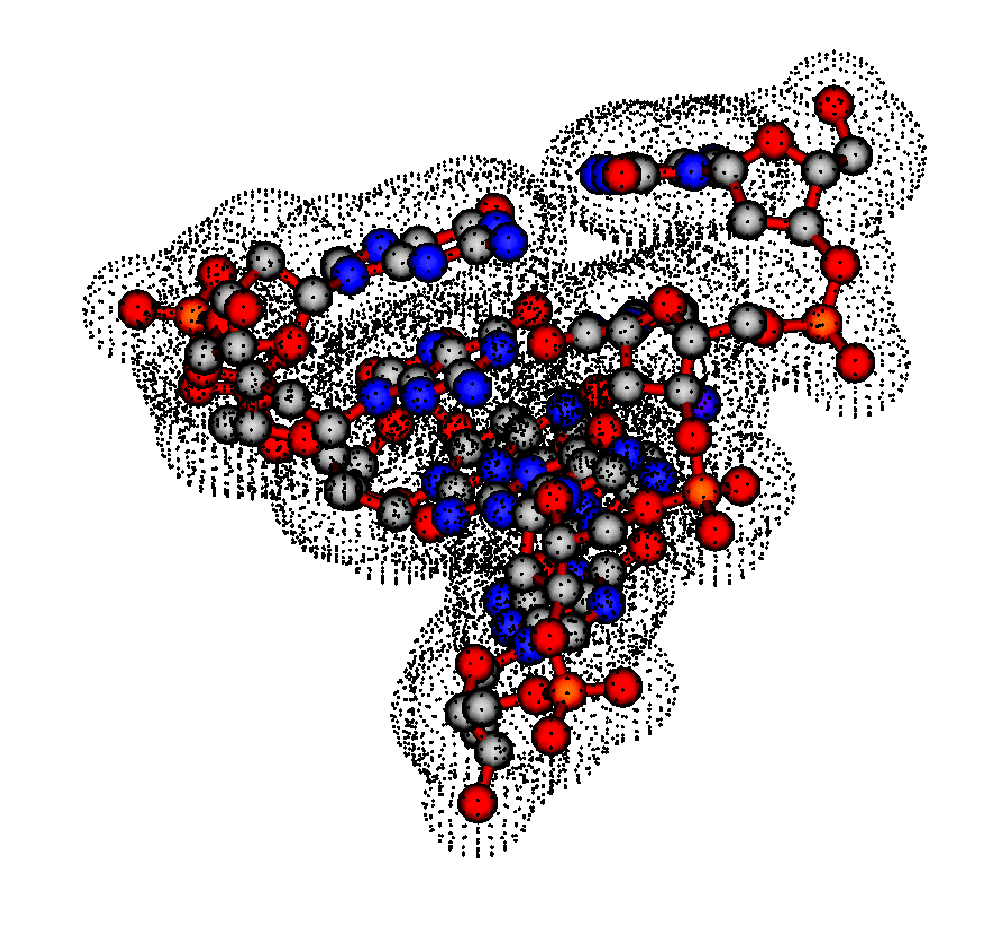}
\includegraphics[width=40mm, height=40mm]{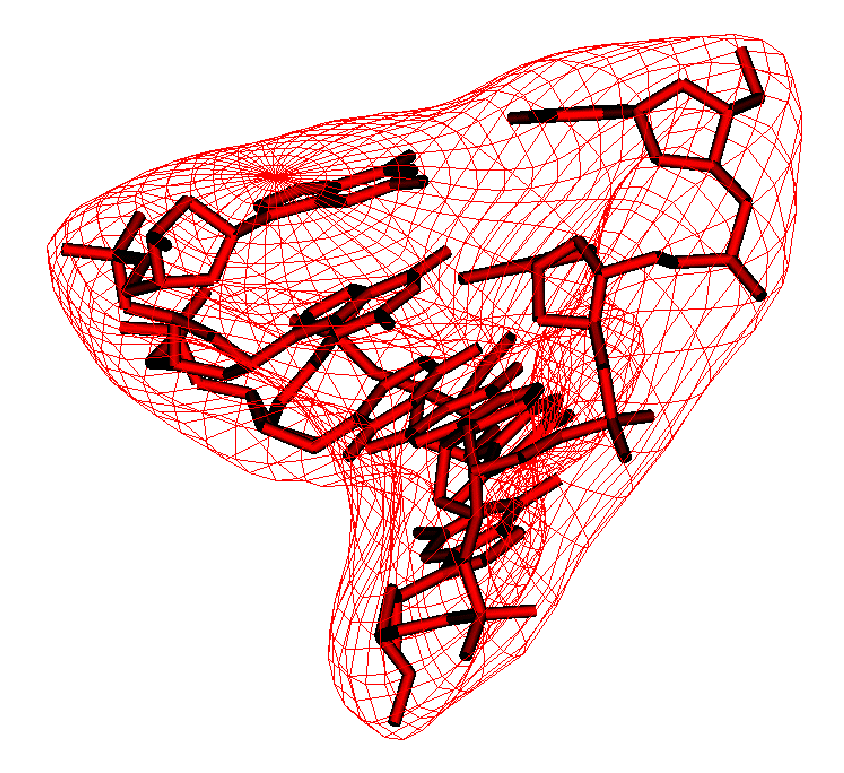} \includegraphics[width=40mm, height=40mm]{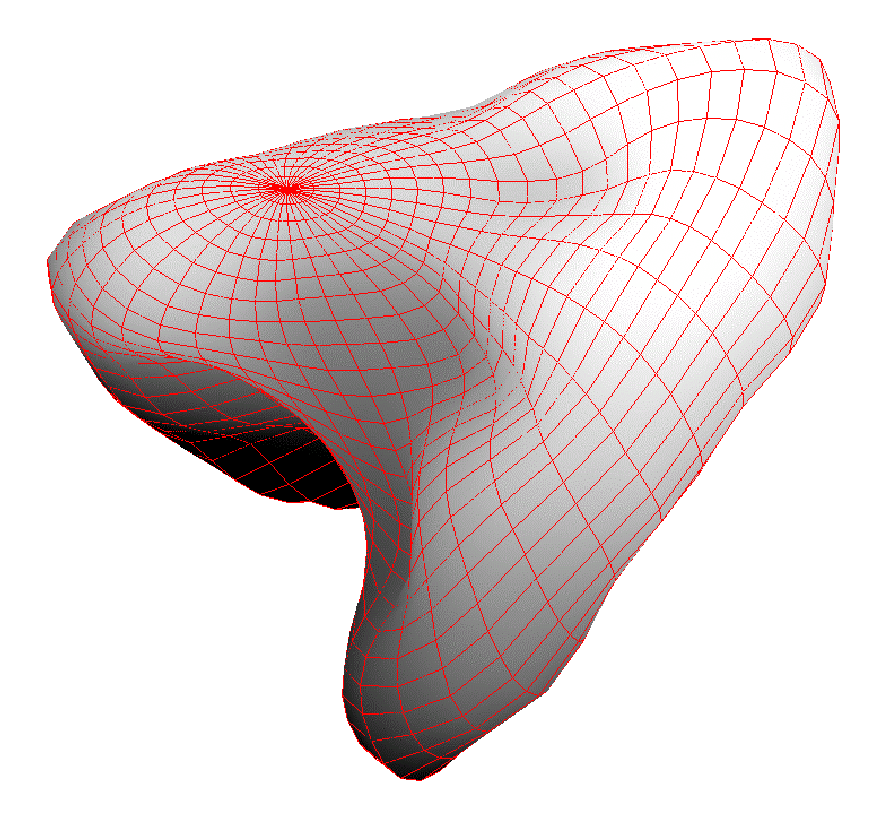}
\caption{Construction of principal surface with spherical topology
for a distribution of points on Van der Waals molecular surface of
a biological molecule \label{Fig7}}
\end{figure}

The third example illustrates modeling of surfaces in 3D. An
interesting challenge is to model \textit{molecular surfaces} of
complex biological molecules like proteins using principal
manifold approach. We extracted the Van-der-Waals molecular
surface, using slightly modified Rasmol source code \cite{Sayle92}
(available from the authors by request) for a simple fragment of
DNA. The topology of the surface is expected to be spherical. We
should notice that since it is impossible to make the lengths of
all edges equal for the sphere-like grid, in the {\textit{elmap
}}package some corrections are performed for edge and rib weights
during the grid initialization (shorter edges are given with
larger weights proportionally and the same for the ribs). As a
result one gets a smooth principal manifold with a spherical
topology approximating rather a complicated set of points. This
also allows us to introduce a global spherical coordinate system
on the molecular surface. The advantage of this method is its
ability to deal not only with star-like shapes as the spherical
harmonic functions approach does (see, for example, \cite{Cai02})
but also to model complex forms with cavities as well as
non-spherical forms. The result of applying the principal manifold
construction by elmap package is shown on Fig.~\ref{Fig7}.

\begin{figure}
\centering{ a)\includegraphics[width=35mm,
height=35mm]{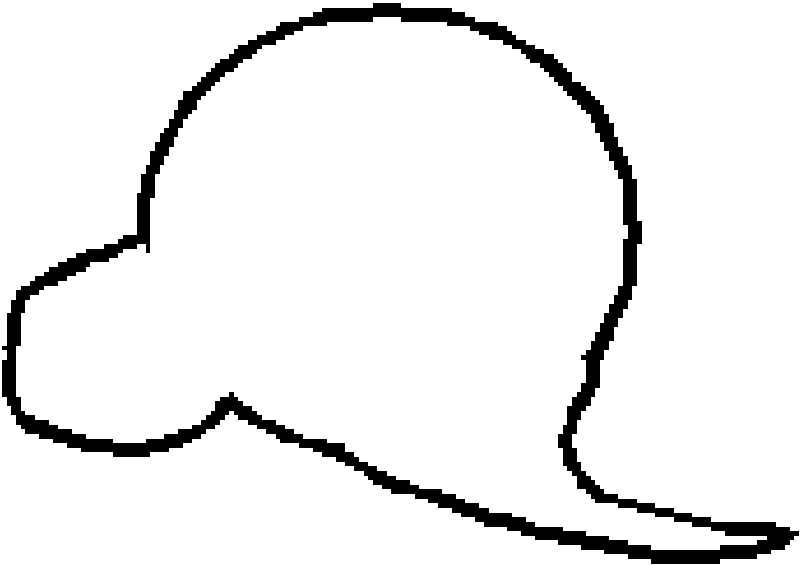} b)\includegraphics[width=35mm,
height=35mm]{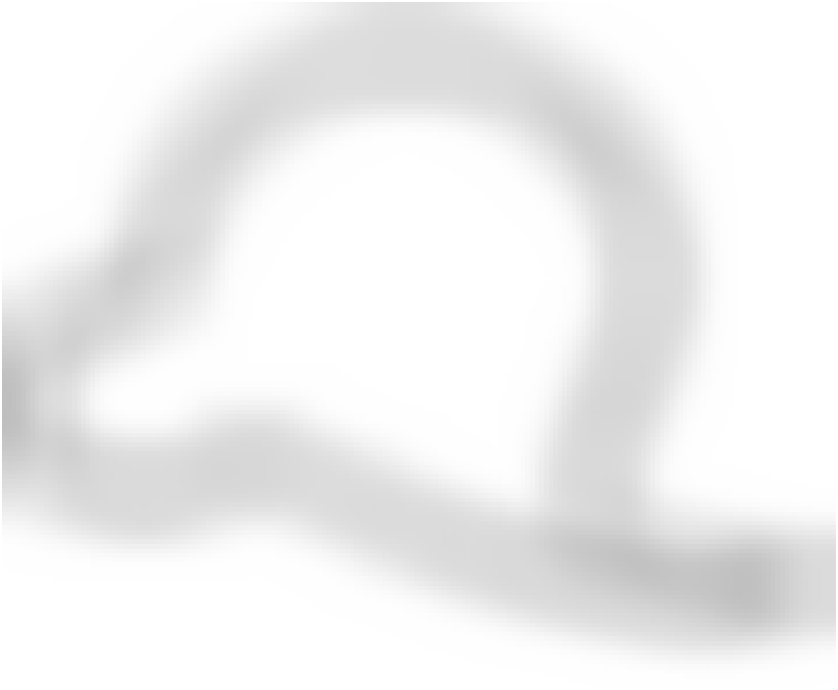} c)\includegraphics[width=35mm,
height=35mm]{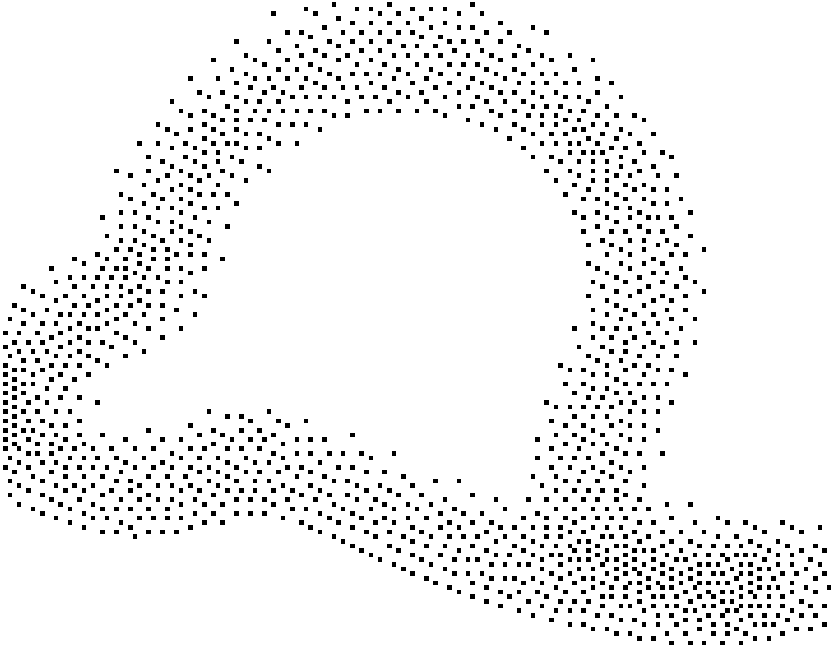} } \\ d)\hspace{5mm}
\includegraphics[width=47mm, height=38mm]{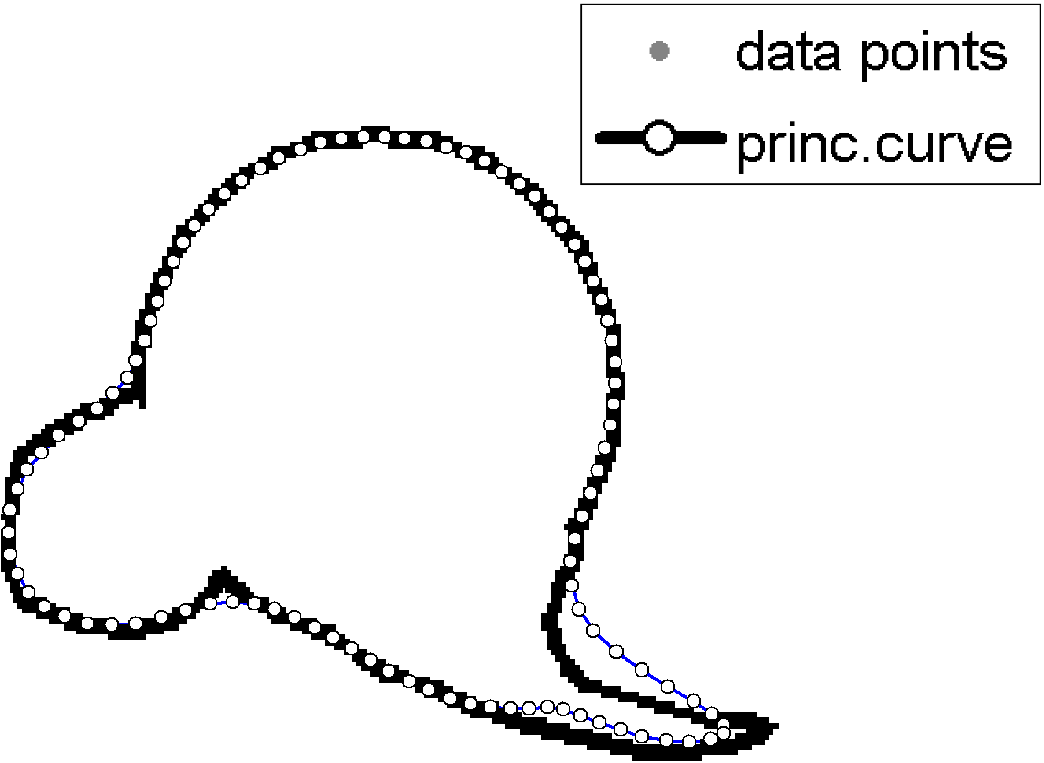}
e)\hspace{5mm}\includegraphics[width=35mm,
height=30mm]{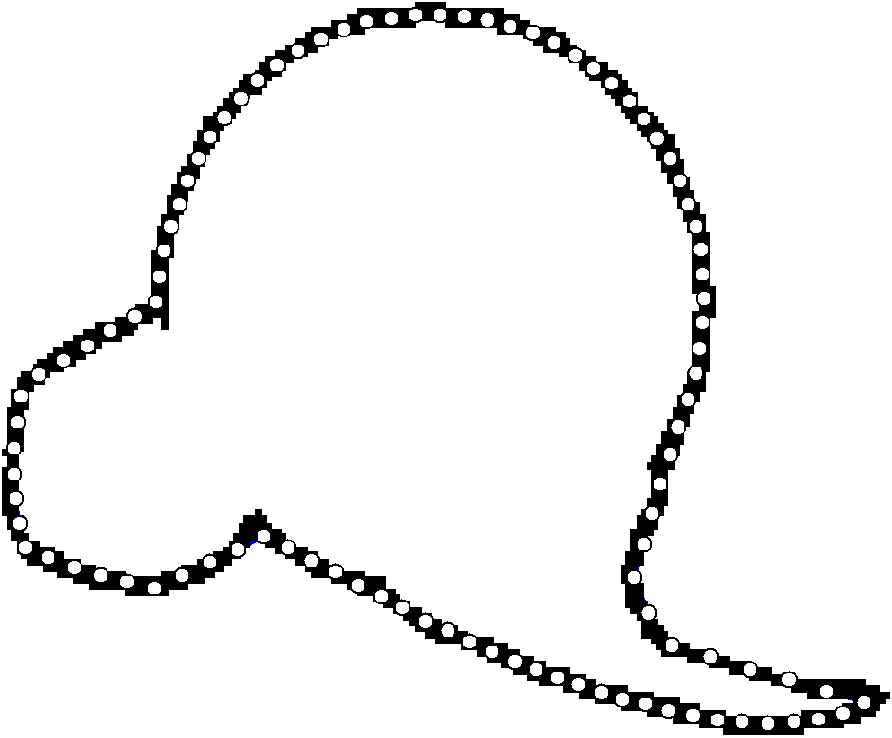} \\ \vspace{6mm}
f)\hspace{8mm}\includegraphics[width=35mm,
height=30mm]{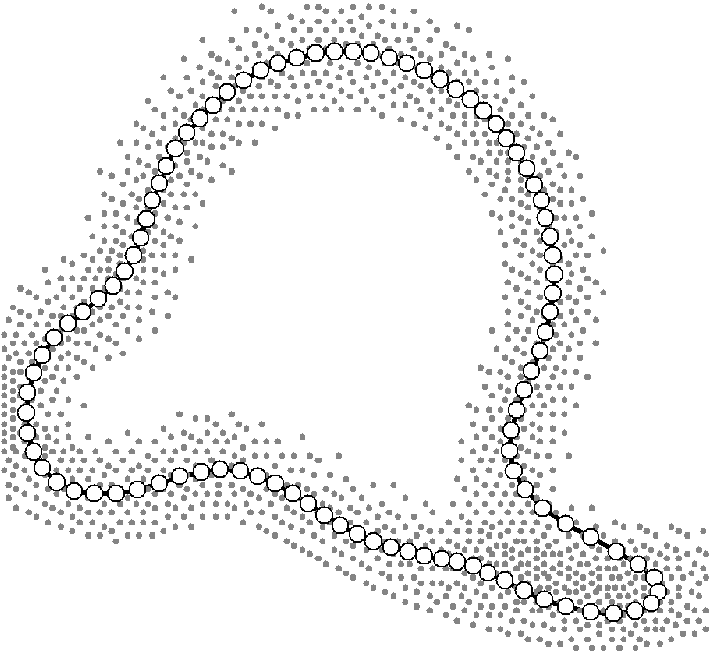} \hspace{13mm}
g)\hspace{8mm}\includegraphics[width=35 mm,
height=30mm]{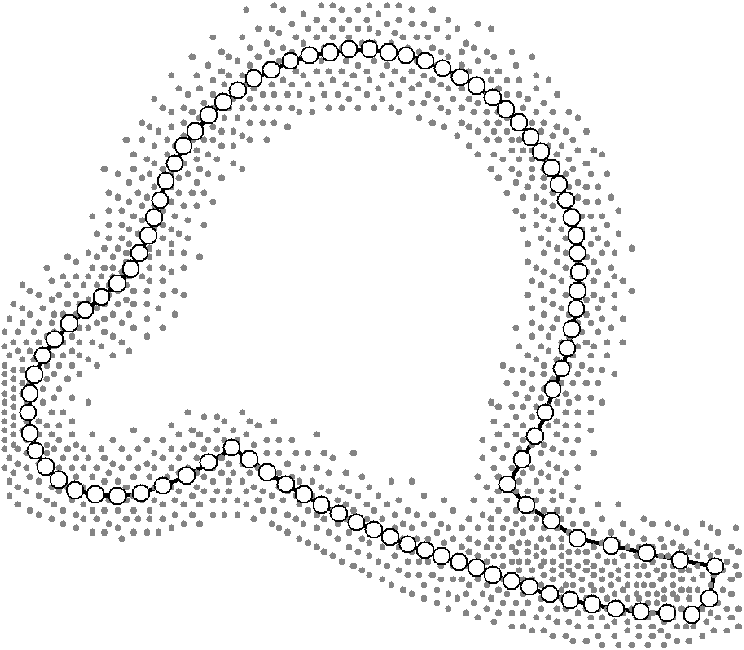} \vspace{6mm} \caption{Contours extraction
with closed principal curve. a) initial countour; b) blurred
contour; c) Floyd-Steinberg error diffusion color image
binarization; d,f) fitting closed principal curve with constant
``elasticity", regions of higher curvature can not be fitted
equally well; e,g) fitting closed principal curve with adaptive
elasticity (``break" adaptation strategy). \label{Fig8} }
\end{figure}

\begin{figure}
\centering{ a)\includegraphics[width=50mm,
height=50mm]{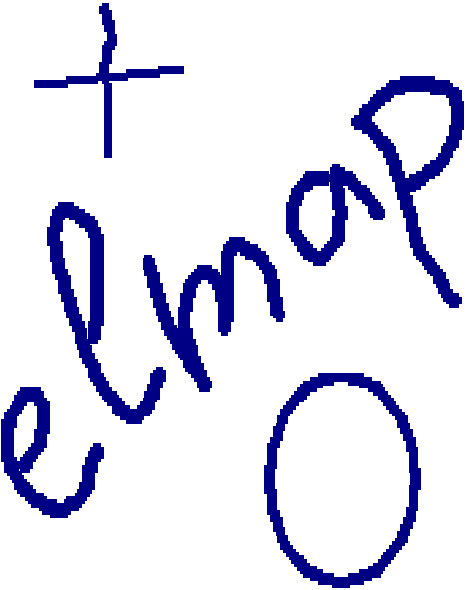} \hspace{5mm}
b)\includegraphics[width=50mm, height=60mm]{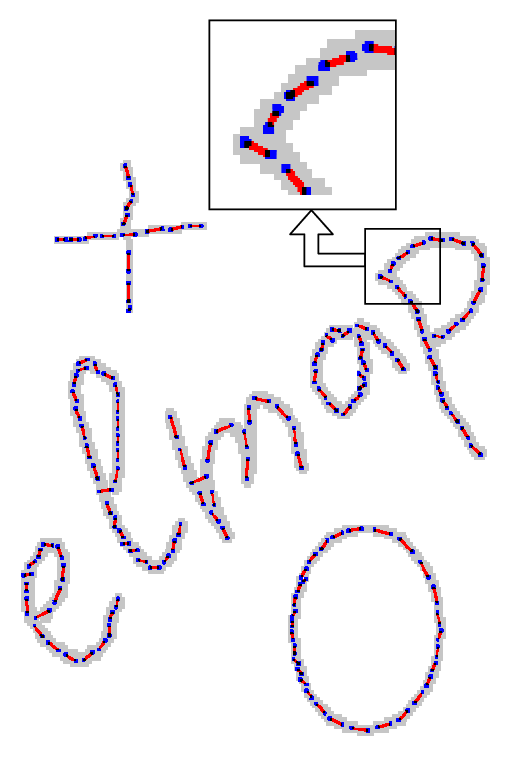} \\
c)\includegraphics[width=50mm, height=60mm]{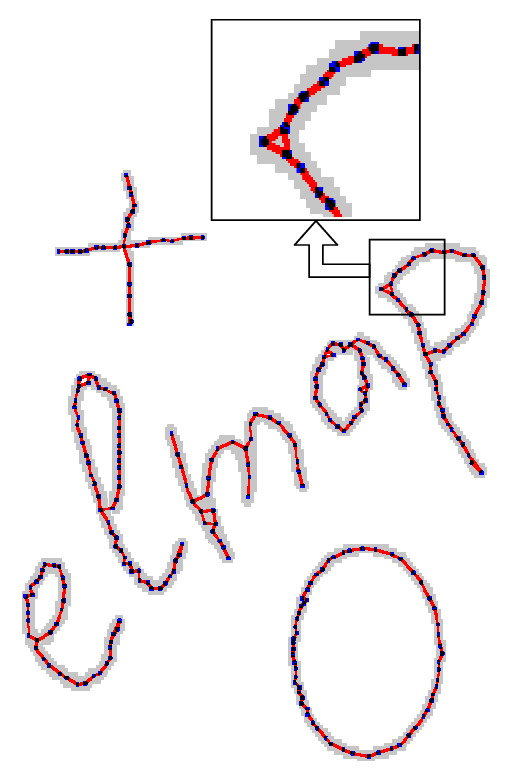}}
\hspace{5mm} d)\includegraphics[width=50mm,
height=60mm]{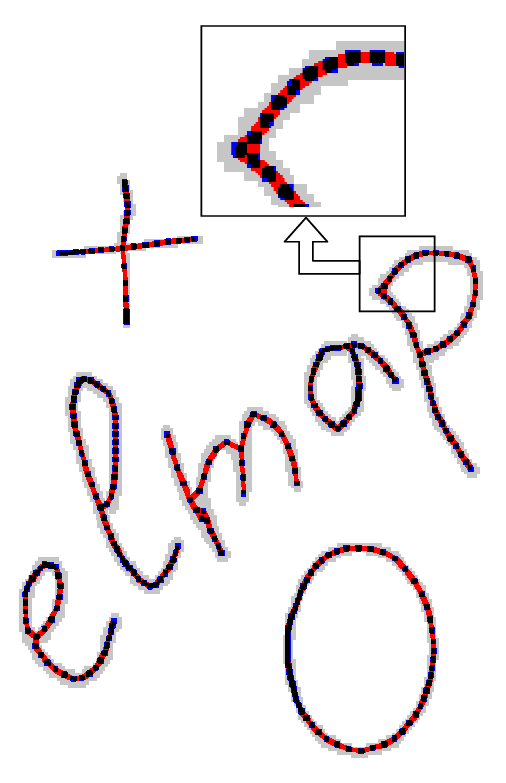} \caption{Skeletonization using principal
curves: a) initial image; b)  calculation of local principal
components; c) connecting the graph; d) graph vertices
optimization with principal components algorithm. \label{Fig9}}
\end{figure}

The forth example demonstrates extracting curvilinear features
from images with the \textit{elmap} package. Fig.~\ref{Fig8}
demonstrates how ``principal graph" strategy is used for contour
extraction. Fig.~\ref{Fig9} shows how ``principal graph'' strategy
is used for skeletonization of hand-written symbols.

\begin{figure}
\centering{ a)\includegraphics[width=35mm,
height=35mm]{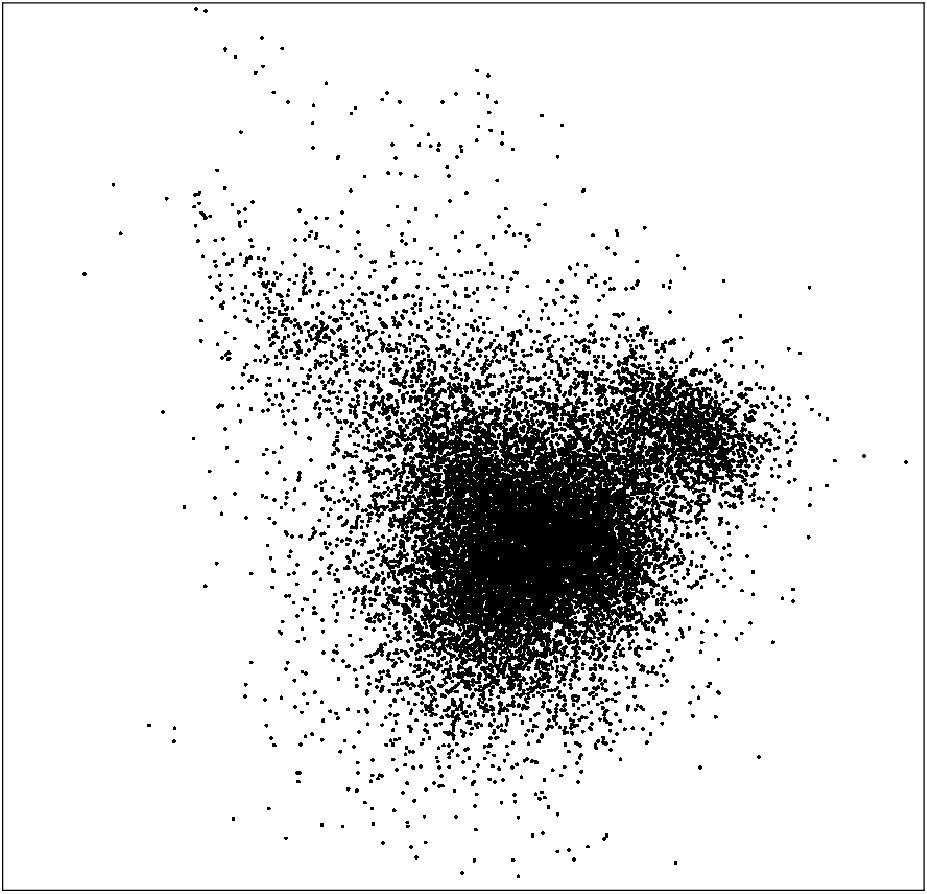} b)\includegraphics[width=35mm,
height=35mm]{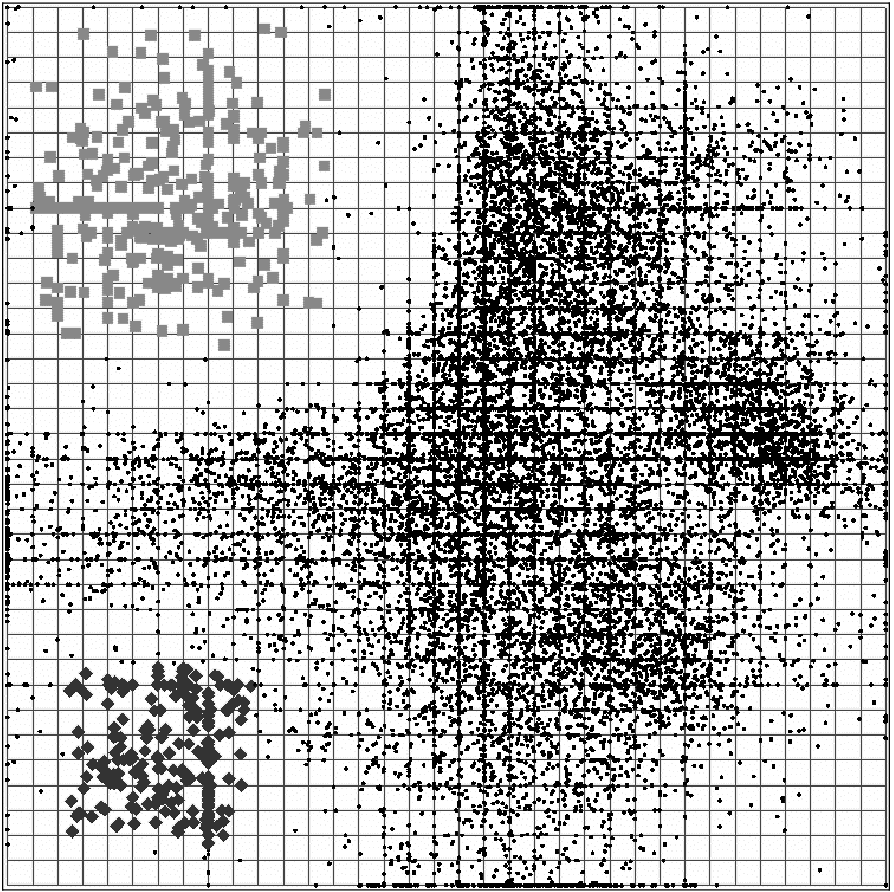} c)\includegraphics[width=35mm,
height=35mm]{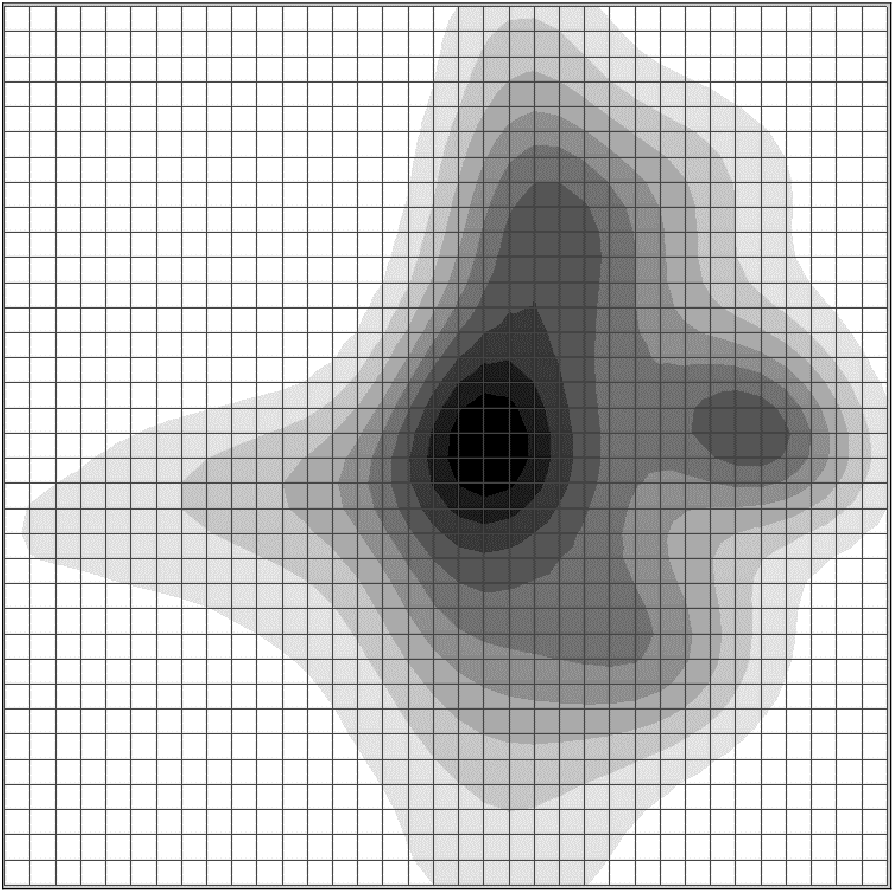} } \caption{Visualization of a big
dataset in 84-dimensional space. a)  PCA view; b) projection onto
the manifold constructed; two strong signals are marked by
changing point sizes/forms; c) principal manifold as a screen for
displaying points density distribution. \label{Fig10}}
\end{figure}

Our final, fifth example illustrates an application of the
principal manifold method in multidimensional \textit{data
visualization} and dimension reduction. As in the case of
molecular surface modeling, we take an example of a dataset from
bioinformatics. The genome of C.eleganse (small worm with only
one-hundred cells) contains approximately 17000 genes, each of
them can be characterized by its codon usage (there are 64 codons,
i.e. triplets of 4 genetic letters, this gives a 64-dimensional
vector of their frequencies), dinucleotide and nucleotide usage
(this gives additional 20 dimensions). The resulting dataset has
17083 points with 84 dimensions. PCA view of the dataset is shown
at Fig.~\ref{Fig10}a. To make noise-filtering, the dataset was
projected first into 25-dimensional space spanned by the first 25
principal vectors. In this space, using our {\textit{elmap}}
package, we constructed a two dimensional principal surface,
approximated by 1296 nodes. The datapoints were projected onto the
manifold by projecting onto the closest point of the manifold (as
proposed above). Using a 3-epoch optimization strategy, provided
in the sample initialization file for the {\textit{elmap}}
package, it takes 300 seconds to do this on a computer with Athlon
1800 MHz processor. The initial mean-square error (MSE), obtained
by a principal plane approximation was 4.59. The resulting
manifold provides MSE about 3.60; what is at 22\% better than
approximation by the principal plane (this value is relatively
big, bearing in mind that we approximate a 25-dimensional
dataset). The resulting image of projections is shown on
Fig.~\ref{Fig10}b. Changing point forms/sizes we marked two
signals that are clearly seen on this plot. More detailed analysis
shows that indeed these two groups of points (genes) have very
special positions in the dataspace (i.e., codons and dinucleotide
compositions) with respect to the main cluster of data. The
principal manifold we constructed can be utilized for displaying
different functions defined in the dataspace. On Fig.~\ref{Fig10}c
visualization of a simple non-parametric estimation of the density
distribution is shown. One can see that in general the non-linear
manifold captures more essential features of the dataset than the
PCA plot.

\section{Method implementation}

In the implementation of the algorithm we used the SparseLib
\cite{Dongarra94} library together with IML++ library to store the
matrix and to solve the system of linear equations. We used the
BLAS kernel provided by the authors of SparseLib without any
platform-specific optimization. This combination showed rather
good performance characteristics, still being easily portable,
i.e. written, using ANSI standards. The \textit{elmap} package
together with a stand-alone data visualization tool
\textit{VidaExpert} are available online \cite{Elmap,VidaExpert}.

\section{Discussion}

We introduced a new algorithmic kernel for calculating grid
approximations for principal manifolds of different topologies and
dimensions. The main advantages of this method are speed and good
performance. The optimization criterion we formulated has a
particularly simple form and natural physical interpretation.
Together with the usual mean square node-to-point distance term
our minimized functional contains two penalizing terms:  $ U^{(E)}
$  and $U^{(R)}$, both quadratic with respect to the grid nodes
positions. As one can see from (\ref{eq2}) and (\ref{eq3}) they
are similar to the sum of squared grid approximations of the first
and second derivatives, in the directions, guided by natural
choice of ribs\footnote{ The differences should be divided by
node-to-node distances in order to be true derivative
approximations, but in this case the quadratic structure of the
term would be violated. We suppose that the grid is regular with
almost equal node-to-node distances, then the dependence of
coefficients \textit{$\lambda $}$_{i}$, \textit{$\mu $}$_{j}  $ on
the total number of nodes contains this factor.} . The $U^{(E)}$
term penalizes the total length (or area, volume) of the principal
manifold and, indirectly, makes the grid regular by penalizing
non-equidistant distribution of nodes along the grid. The
$U^{(R)}$ term is a smoothing factor. It penalizes the
nonlinearity of the ribs embedding into the Euclidean space. This
term is quadratic, it gives us some benefits in comparison with
the cosine function as in the algorithm of K\'{e}gl \cite{Kegl99},
for example.

Attractive characteristics of the method such as its universality,
speed and inherited parallelism open new fields to the
applications of principal manifolds, especially for the analysis
of huge datasets with hundreds of thousands of points with
dimensionality of the order of hundreds. The algorithm we
described with its C++ implementation provide a way to construct a
principal manifold for these datasets approximated by a number of
nodes of the order of 10000 in a reasonable time.

In applications of principal manifolds to 3D-surface modeling, one
can find similar ``physics-based'' new methods in surface modeling
in computer graphics (see, for example \cite{Mandal00,Xie02}). The
method of constructing the elastic energy functional considered
here can be compared with approach described in \cite{Xie02}. Our
functional contains only restricted subset of elastic energies
proposed there; we utilize such a ``physics-based'' model, which
allows quadratic description, thus leads to quadratic optimization
problem. In this way we significantly speed-up the optimization
step. Also one can consider use of physically realistic energy
functionals and pre-defined equilibrium forms as described above.
In general, our point is to construct computationally effective
approximation method rather than closely imitate realistic
behavior (though it is also possible): this is particularly true
for multidimensional applications where the notion of ``physical
realism'' does not make sense.

One important application of principal manifolds is dimension
reduction and data visualization. In this field they compete with
multidimensional scaling methods and the recently introduced
advanced algorithms of dimension reduction, such as locally linear
embedding (LLE) \cite{Roweis00} and ISOMAP \cite{Tenenbaum00}
algorithms. The difference between the two approaches is that the
later ones seek new point coordinates directly and do not use any
intermediate geometrical objects. This has several advantages, in
particular that a) there is a unique solution for the problem (the
methods are not iterative in their nature, there is no problem of
grid initialization) and b) there is no problem of choosing a good
way to project points onto a non-linear manifold. Another
advantage is that the methods are not limited by several first
dimensions in dimension reduction (it is difficult in practice to
manipulate non-linear manifolds of dimension more than three).

Principal manifold can serve as a non-linear low-dimensional
screen to project data. It gives additional benefits  to users.
First, the manifold approximates data and can be used itself,
without applying projection, to visualize different functions
defined in data space (for example, density estimation). Also the
manifold as an intermediate, ``fixing'' the structure of a
learning dataset, can be used in visualization of data points that
were not used in the learning process, for example, for
visualization of dataflow ``on the fly''. Constructing manifolds
does not use a point-to-point distance matrix that is particularly
useful for large datasets. Also using principal manifolds is
expected to be more robust to additive noise than the methods
based on the local properties of point-to-point distances. To
conclude this short comparison, LLE and ISOMAP methods are more
suitable if the low-dimensional structure in multidimensional data
space is complicated but is expected to exist, and if the data
points are situated rather tightly on it. Principal manifolds are
more applicable for the visualization of real-life noisy
observations, appearing in economics, biology, medicine and other
sciences, and for constructing data screens showing not only the
data but different related functions defined in data space.

\section{Appendix. Constructing the sparse matrix}

Matrix (6) has $p^{2}$ elements (where $p$ is a number of grid
nodes), but for typical grids only $kp$ of them are non-zero,
where $k \ll p $ . Here we provide a simple procedure to fill only
non-zero elements of the matrix, thus, define its sparse
structure.

For the  $ e^{jk} $  matrix:
\begin{enumerate}
\item{All $e^{jk}$ values are initialized by zero;}
\item{If for an edge $E^{i}$ with weight \textit{$\lambda $}$_{i}$,
the beginning node is $y^{k1} $ and the ending node is $y^{k2}$,
then we update the $e^{jk}$ values:

\[
e^{k_{1} k_{1}}  = e^{k_{1} k_{1}}  + \lambda _{i} , e^{k_{2} k_{2}}
= e^{k_{2} k_{2}}  + \lambda _{i} , e^{k_{1} k_{2}}  = e^{k_{1}
k_{2}}  - \lambda _{i} , e^{k_{2} k_{1}} = e^{k_{2} k_{1}}  -
\lambda _{i} .
\]
} \item{Steps 1-2 are repeated for every edge.}
\end{enumerate}
For the $r^{jk}$ matrix:
\begin{enumerate}
\item{All $r^{jk}$ values are initialized by zeros;}
\item{If for a rib $R^{i}$ with weight \textit{$\mu
$}$_{i}$, the beginning node is $y^{k1}$, the middle node is
$y^{k2}$ and the ending node is $y^{k3}$, then we update the
$r^{jk}$ values:

\[
r^{k_{1} k_{1}}  = r^{k_{1} k_{1}}  + \mu _{i} , r^{k_{2} k_{2}}  =
r^{k_{2} k_{2}}  + 4\mu _{i} , r^{k_{3} k_{3}}  = r^{k_{3} k_{3}}  +
\mu _{i}
\]

\[
r^{k_{1} k_{2}}  = r^{k_{1} k_{2}}  - 2\mu _{i} , r^{k_{2} k_{1}}  =
r^{k_{2} k_{1}}  - 2\mu _{i} ,
\]

\[
r^{k_{2} k_{3}}  = r^{k_{2} k_{3}}  - 2\mu _{i} , r^{k_{3} k_{2}}  =
r^{k_{3} k_{2}}  - 2\mu _{i} ,
\]

\[
r^{k_{1} k_{3}}  = r^{k_{1} k_{3}}  + \mu _{i} , r^{k_{3} k_{1}}  =
r^{k_{3} k_{1}}  + \mu _{i} .
\]
}
\item{Steps 1-2 are repeated for every rib.}
\end{enumerate}

\bigskip

\noindent Alexander Gorban

\noindent University of Leicester

\noindent University Road

\noindent Leicester, LE1 7RH

\noindent UK

\noindent ag153@le.ac.uk

\bigskip

\noindent Andrey Zinovyev

\noindent Institut Curie

\noindent 26, rue d'Ulm

\noindent Paris, 75248

\noindent France

\noindent zinovyev@ihes.fr

\end{document}